\documentclass{aa}  

\usepackage{graphicx}
\usepackage{txfonts}
\usepackage{hyperref}
\hypersetup{colorlinks,
  linkcolor=blue,
  filecolor=blue,
  urlcolor=blue,
  citecolor=blue}

\begin{document}

   \title{The formation and evolution of Supermassive disks in IllustrisTNG}
   \author{Diego Pallero
          \inst{1}
          \and
          Gaspar Galaz\inst{2}
          \and
          Patricia B. Tissera\inst{2,3,4}
          \and
          Facundo A. G\'omez\inst{5}
          \and
          Antonela Monachesi\inst{5}
          \and
          Cristobal Sif\'on\inst{6}
          \and
          Brian Tapia-Contreras\inst{2}
          }

   \institute{Departamento de F\'isica, Universidad T\'ecnica Federico Santa Mar\'ia, Avenida Espa\~na 1680, Valpara\'iso, Chile\\
              \email{diego.pallero@usm.cl}
         \and
Instituto de Astrofísica, Pontificia Universidad Católica de Chile, Avenida Vicuña Mackenna 4860, 7820436, Macul, Santiago, Chile
          \and
          Millenium Nucleus ERIS, Chile  
          \and
          Centro de Astro-Ingeniería, Pontificia Universidad Católica de Chile, Av. Vicuña Mackenna 4860, Santiago, Chile
          \and
          Departamento de Astronomía, Universidad de La Serena, Av. Ra\'ul Bitr\'an 1305, La Serena, Chile
          \and
          Instituto de Física, Pontificia Universidad Católica de Valparaíso, Casilla 4059, Valparaíso, Chile
          }

   \date{Received September 15, 1996; accepted March 16, 1997}

  \abstract
   {Supermassive disks are outstanding galaxies whose formation and evolution are still poorly understood. They comprise a large variety of objects, from large low-surface brightness galaxies as Malin-1 to the most spectacular superluminous spirals. However, we still do not know the physical mechanisms behind its formation, and whether they will be long-lived objects or whether their mass could destroy them in time.}
   {We aim to investigate the formation and evolution of supermassive disks in the magnetohydrodynamical state-of-the-art simulation IllustrisTNG-100. We aim to characterize the main formation mechanisms behind these galaxies and to understand the role played by the environment in their formation.}
   {Supermassive disks were identified using morpho-kinematic criteria, based on a relation between the spin-parameter $\lambda_R$ and its ellipticity ($\varepsilon$) with two different thresholds.
   We defined supermassive disks as galaxies with $\lambda / \sqrt{\varepsilon} \geq 0.31$ or 0.71, and with stellar mass log$_{10}M_\star/M_\odot > 10^{11}$. We studied the color, merging history, AGN history, and environment in which these galaxies reside. Additionally, we studied galaxies individually to check how they formed.}
   {Supermassive disk galaxies typically experience a quiescent merging history, with $48\%$ experiencing no significant mergers at $z \leq 1$. Their stellar mass growth is driven mainly by star formation, unlike spheroidal galaxies, which require a significant number of mergers to form. Moreover, the mergers experienced by disk galaxies are generally rich in gas content, irrespective of whether they are minor or major events. Supermassive disks exist across various environments, from isolation to clusters, with $\sim 60\%$ inhabiting in isolation or low-mass groups, $\sim 25\%$ residing in massive groups, and $\sim 15\%$ residing within galaxy clusters. When studying the evolution of supermassive disks selected at $z=0.5$, we show that when they gain sufficient mass, the probability of them maintaining their disk-like structure up to $z=0$ is relatively high ($\sim 60\%$). Lastly, while AGN significantly influences the regulation of star formation in galaxies, it does not directly alter their morphological structure.}
   {}


   \maketitle
%

\section{Introduction}

Galaxies in the Universe can take various shapes and forms. Since the first half of the XX century, classification schemes have been developed to organize them by their morphology, the most famous being the Hubble Sequence \citep[][]{Hubble26}.
\begin{figure*}
\centering
\includegraphics[width=\textwidth]{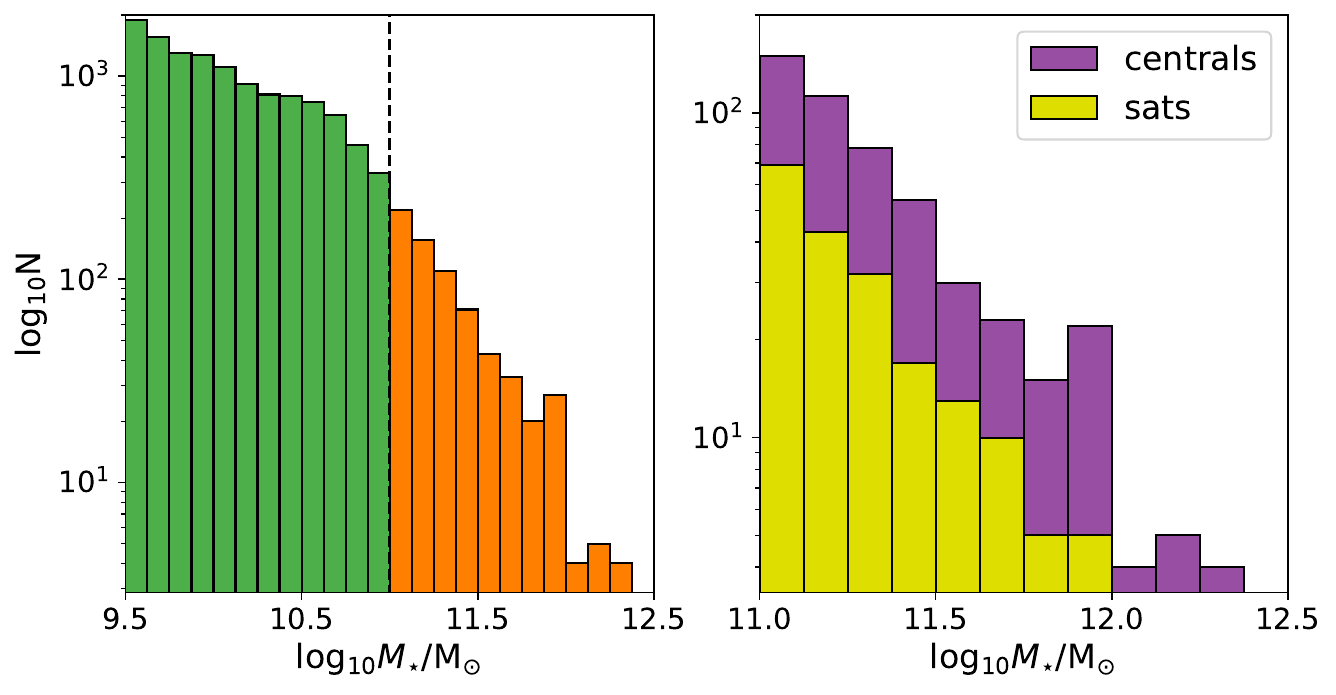}
\caption{Left: Stellar-mass distribution for all galaxies in TNG100. The dashed line shows the limit log for which we consider galaxies as ``massive''. The orange bars comprise the core of our study sample. Right: Satellite (yellow)—central (purple) separation for galaxies with log$_{10}M_\star/M_\odot > 10^{11}$. Massive galaxies are predominantly central, even though a non-negligible fraction are satellites.}

\label{fig:galaxy_dist}
\end{figure*}

Nowadays, it is well understood that morphology is closely related to some more fundamental properties of galaxies. Elliptical/early-type galaxies tend to exhibit lower star formation rates, an older population, and, consequently, redder colors. On the other hand, spiral/late-type galaxies show higher star formation rates, a population dominated by young stars and, consequently, showing bluer colors \citep[][]{Morgan57,Gomez03,Kauffmann04,Blanton09}. 

A fundamental pending question is how the morphology of galaxies evolves in the transition zone between late- and early-type galaxies. This transition zone has been called "Green Valley", given the vast and relatively flat shape of the region occupied by galaxies in the color-magnitude diagram\citep{Wyder07,Schiminovich07,Martin07}.
Morphological transformations are likely to occur throughout a galaxy's history. Early observations of the Universe have indicated a prevalence of luminous disk-like galaxies \citep{Kolesnikov24}. Later on, with the arrival of deep observations, it has been established that most nearby dwarf galaxies are spheroidal \citep{Bernardi03,Wuyts11,Conselice14,Shibuya15}.
Given the hierarchical paradigm of structure formation \citep{White&Rees78}, these transformations in morphology from rotationally supported disks to dispersion-dominated spheroids are widely explained by galaxy mergers \citep[][]{Toomre72,Barnes92,Bournaud07,Martin18, Montaguth23,Montaguth24}{}{}. 
This process becomes crucial towards the high-mass end of the galaxy mass function \citep{Faber07,McIntosh08,Cattaneo11,Kaviraj17}.
After the knee of the galaxy mass function (log$_{10}M_\star \geq 10.8$[M$_\odot$]), mergers are thought to be needed for galaxies to achieve such high stellar masses. In this sense, spheroidal galaxies are expected to dominate the high-mass regime.

Within this context, \citet[][]{Quilley22}{}{} revisited the relation between morphology and colour using the extensive morphological classification from the EFIGI (“Extraction de Formes Id\'ealis\'ees de Galaxies en Imagerie”) galaxies \citep[][]{deLapparent11}{}{}, a subsample of MorCat (``Morphology Catalogue'') from the Sloan Digital Sky Survey (SDSS) Data Release 8 \citep[DR8;][]{Aihara11}{}{}. This work proposes that the Hubble Sequence can be understood as an inverse sequence of galaxy evolution. In this sense, internal dynamical processes, such as a dominant presence of bars and strong flocculent arms in galaxies at the high-mass end, suggest that internal dynamics of massive disk galaxies, probably triggered by fly-by encounters or minor mergers, can trigger the bulge growth in the green valley. 
This systematic bulge growth across the green valley highlights the region's transitional nature while excluding a predominantly quick transit due to rapid quenching. However, it is worth noting that this result may change when considering the environment in which galaxies reside \citep[e.g.][]{Pallero20}.

A key result highlighted in \citet{Quilley22} is the lack of massive disks compared to elliptical galaxies. Specifically, they did not find disks more massive than log$_{10}M_\star \geq 11.8$[M$_\odot$], proposing an upper limit for disk galaxies to exist. 
These results are in agreement with what was found by \citet[][]{Cappellari16}{}{}, which shows a critical mass of log$_{10}M_\star > 11.3$[M$_\odot$] splitting between fast (disks) and slow (spheroids) rotators.
This result was also explored in \citet[][]{Jackson20}{}{} using galaxies from the Horizon-AGN cosmological hydrodynamical simulation \citep[][]{Dubois13}{}{}.
They found two main formation mechanisms for massive disks (log$_{10}M_\star > 11.4$[M$_\odot$]): a rejuvenation scenario and a quiescent merging history. 
The primary one is the rejuvenation scenario, which accounts for 70$\%$ of their sample of massive disks. In this case, spheroidal galaxies experience a significant gas-rich merger (mass ratio $\geq$ 1:10), recreating the disk from a rotationally supported galaxy. Mergers create a new rotational stellar component, leaving a massive disk. 
The secondary channel accounts for the remaining 30$\%$ of their sample. They show that these galaxies experience a quiescent merger history and show a disk morphology throughout their history.
It should be noted that the fraction of massive disks that these authors find accounts for $\sim 11\%$ of massive galaxies within their simulations. Later on, \citet[][]{Jackson22}{}{} found a similar fraction of massive disks in a sample of observed galaxies from the SDSS using the MPA-JHU value-added catalogue\footnote{\url{https://wwwmpa.mpa-garching.mpg.de/SDSS/DR7/}}. 
{Although significant efforts have been made to study these types of objects, supermassive disks remain scarce, with less than 200 objects detected to this day \citet{Ogle19, Jackson22}. We expect this knowledge gap to be filled with the next-generation wide-field surveys such as EUCLID \citep{EUCLID} and the Vera C. Rubin Observatory's Legacy Survey of Space and Time \citep[LSST][]{LSST}.

In this work,  we revisit the formation mechanisms of extremely massive disk galaxies with the state-of-the-art IllustrisTNG simulations. We aim to investigate the likelihood of these galaxies forming, the various formation mechanisms behind the formation of massive disks, and to understand whether these massive disks are long-lived structures or merely a transitional phase during the transformation from disks to spheroids.
The article is organized as follows. In Section \ref{sec:data}, we summarize the relevant aspects of IllustrisTNG. We define our galaxy sample and the procedures performed to measure the morpho-kinematical properties of galaxies. In Section \ref{sec:results}, we show the disk fraction in the galaxy-mass function and the positions that massive disks occupy in the colour-mass diagram, providing crucial insights into the formation of these galaxies. We also quantify the significance of mergers in building up the stellar components of massive disks. In Section \ref{sec:discussion}, we investigate whether massive disks are long-lived structures and the different environments in which they can exist, shedding light on the longevity and diversity of galaxies. In addition, we compare the formation mechanisms identified in this work with those previously reported in the literature, thereby further enhancing our understanding of galaxy evolution and its implications for future research.
Finally, in Section \ref{sec:summary}, we summarize our results and conclusions.

\section{Data}
\label{sec:data}
\subsection{IllustrisTNG}

IllustrisTNG \citep[The Next Generation, TNG hereafter;][]{Marinacci18,Naiman18, Nelson18, Pillepich18,Springel18}\footnote{\url{https://www.tng-project.org/}} is the natural successor of the Illustris\footnote{\url{https://www.illustris-project.org/}} project \citep[][]{Genel14, Vogelsberger14b,Vogelsberger14a,Nelson15,Sijacki15}. It corresponds to a suite of magneto-hydrodynamical simulations with a comprehensive galaxy formation model, which includes prescriptions for star formation, stellar evolution, chemical enrichment, stellar feedback, galactic outflows, black-hole formation, AGN feedback and metal cooling \citep[see][for a detailed description]{Pillepich18a}{}{}.  

This suite offers three different simulation volumes for TNG50, TNG100 and TNG300, with cosmological volumes of $(51.7\, \mathrm{cMpc})^{3}$, $(106.5\, \mathrm{cMpc})^{3}$ and $(302.6\, \mathrm{cMpc})^{3}$ respectively.
This work will only use the TNG100, the intermediate resolution and size simulation. 
This simulation presents the best compromise between a large and representative sample of different galaxies and environments. It reaches the high-mass end of the mass function with a good sample of galaxies while maintaining an excellent resolution to measure the kinematics and morphology.
For this simulation, the mean baryonic particle resolution is $m_{bar} \sim 1.4\times10^{6}{\rm M_\odot}$(gas and stars) while having a dark matter particle resolution of $m_{\rm DM} \sim 7.5\times10^{6}{\rm M_\odot}$.

The simulations were performed using the \textsc{arepo} code \citep[][]{Springel10}{}{}, a gravitational solver for Poisson's equations, using a tree-particle-mesh algorithm. Additionally, it solves equations for magnetohydrodynamics using a Voronoi tessellation method within the simulation domain \citep[][]{Pakmor11}{}{}.
All simulations start at z=127 and adopt a flat $\Lambda$ cold dark matter ($\Lambda$CDM cosmology, whose parameters were calibrated with the data obtained by the Planck Mission \citep[][]{Planck16}{}{}: Energy density $\Omega_\Lambda = 0.6911$, matter density $\Omega_m = 0.3089$, baryonic density $\Omega_b = 0.0486$, a Hubble constant $H_0 = 100h {\rm km s^{-1} Mpc^{-1}}$ with $h = 0.6774$, a $\sigma_8 = 0.8159$ and a spectral index $n_s = 0.9667$.

Several observable have been tested within the simulation and show remarkable agreement with observations, such as the color bimodality for galaxies at $z=0$ \citep[][]{Nelson18}{}{}, the evolution of the mass-metallicity relation \citep[][]{Torrey18}{}{}, the mass-size relation for quiescent and star-forming galaxies \citep[][]{Genel18}{}{} and the metal distribution within the Intra-cluster medium (ICM) \citep[][]{Vogelsberger18}{}{}.
Additionally, within the publicly available catalogues, the simulations include fluxes for galaxies in several bands. Here, we briefly summarise the most important aspects of their
computation, but we refer the reader to \citet{Nelson18} for a more detailed explanation. First, each star particle is treated as a single-burst simple stellar population (SSP) with recorded formation time, metallicity, and initial mass. Their spectral properties were modeled using the FSPS code \citep{Conroy09}, with Padova isochrones, the MILES stellar library, and a Chabrier IMF \citep{Chabrier03}. The isochrone grid spans 22 metallicity and 94 age bins (in log space), and SSP spectra are convolved with SDSS filter responses (u, g, r, i, z at airmass 1.3), including default dust and nebular emission models. 
Band magnitudes for each star particle are computed via bicubic interpolation on the isochrone grid and scaled by initial mass. Total galaxy magnitudes are obtained by summing all subhalo star particles.

For these reasons, TNG100 is an excellent laboratory for studying galaxy evolution and understanding the reasons behind morphological transformations in various environments.

\subsection{Galaxy Sample}

We will rely on the Group Catalogues available in the public TNG database to select galaxies in the simulations. The main properties of haloes and subhaloes are determined using the friends-of-friends (FOF) and \textsc{subfind} algorithms \citep{Springel01, Dolag09} for the 100 snapshots spanning a redshift range between $0 < z < 20$. First, a standard FOF group finder with a linking length $b=0.2$ is used to identify haloes. Then, \textsc{subfind} recognizes and hierarchically characterizes the gravitationally bound substructures. The subhalo catalogue includes centrals and satellites, where the central matches the position of FOF haloes, defined as the position of the dark matter particle with the minimum gravitational potential. Central galaxies may host one or more subhaloes. These subhaloes may or may not contain stellar particles.

This article defines galaxies as all those objects in the subhalo catalogue with a stellar content larger than $M_{\star} \geq 10^{9.5}$ M$_\odot$. Consequently, we ensure a minimum resolution of approximately 1000 stellar particles per object.
Figure \ref{fig:galaxy_dist} shows the stellar-mass distribution for all the galaxies in TNG-100. The simulation comprises 12535 objects at $z=0$ with a stellar mass larger than $M_{\star} \geq 10^{9.5}$ M$\odot$ (represented by the green bars). Among these objects, 694 have a stellar mass larger than $M{\star} \geq 10^{11}$ M$_\odot$ (depicted by the orange bars) and constitute our primary sample for the study.
Given the nature of our study, for each galaxy in the primary sample, we have at least $\sim 10^5$ stellar particles per galaxy, enabling us to measure the kinematics of our objects with reasonable confidence.

As mentioned earlier, central galaxies are those designated as the dominant galaxies within each FOF halo, while satellites are defined as subhaloes hosted by the central halo.
In the right panel of Figure \ref{fig:galaxy_dist}, the central-satellite distribution is shown exclusively for the massive galaxies in our sample. As expected, on average, satellite galaxies tend to be slightly less massive and constitute 27.9$\%$ of the entire sample of supermassive galaxies, indicating that central galaxies primarily dominate the high-mass bin.
The magnitudes, stellar masses, $M_{200}$, and $R_{200}$ used in this study are publicly available in the IllustrisTNG database.
Additionally, to split the sample between disk and spheroids, we performed a morpho-kinematical analysis of each galaxy in our sample, as described below.

\begin{figure}
\centering
\includegraphics[width=0.5\textwidth]{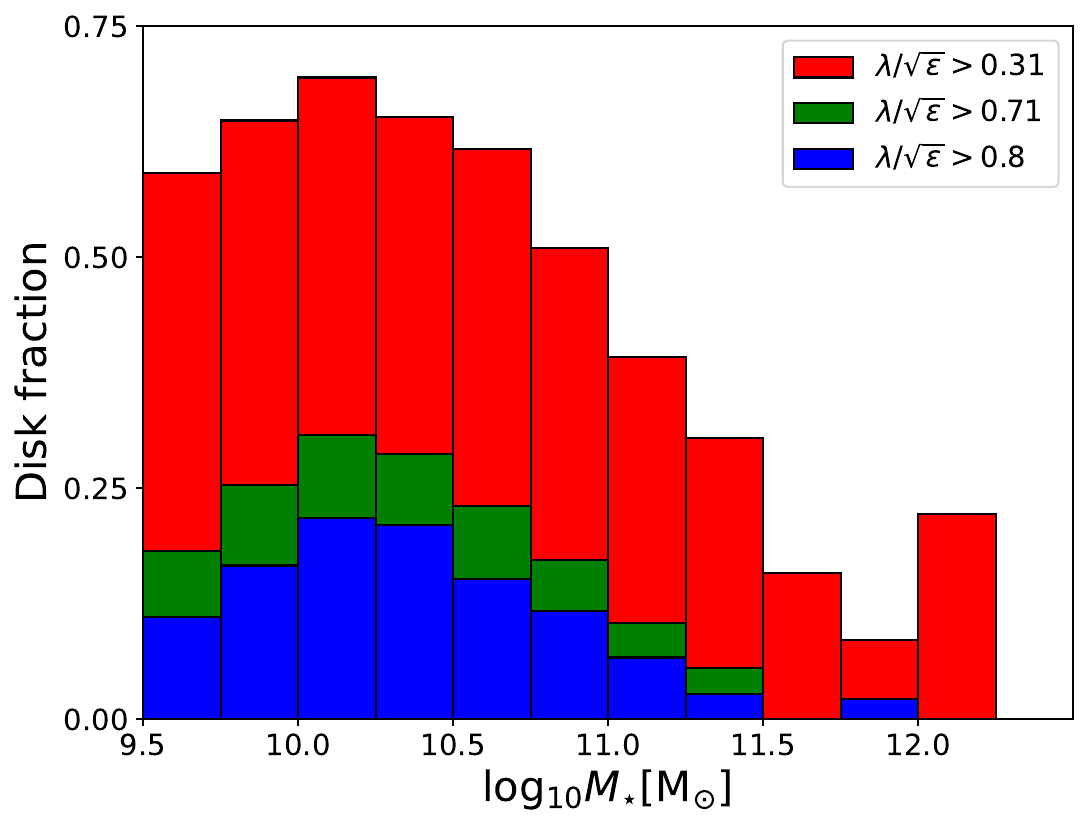}
\caption{Disk fraction for the IllustrisTNG simulation, measured with three different threshold criteria, $\lambda/\sqrt{\varepsilon} = 0.31; 0.71; 0.8$ in red, green and blue, respectively. As we move towards larger stellar mass, the fraction of disks decreases smoothly, with a significant decrease at log$_{10}M_\star \geq 11$[M$_\odot$], reaching values $\sim$ 0 for log$_{10}M_\star \geq 12$[M$_\odot$] regardless of the selected criterion. The results for $\lambda/\sqrt{\varepsilon} =$ 0.71 and 0.81 show little disagreement.}

\label{fig:disk_fraction}
\end{figure}

\begin{figure*}
\centering
\includegraphics[width=\textwidth]{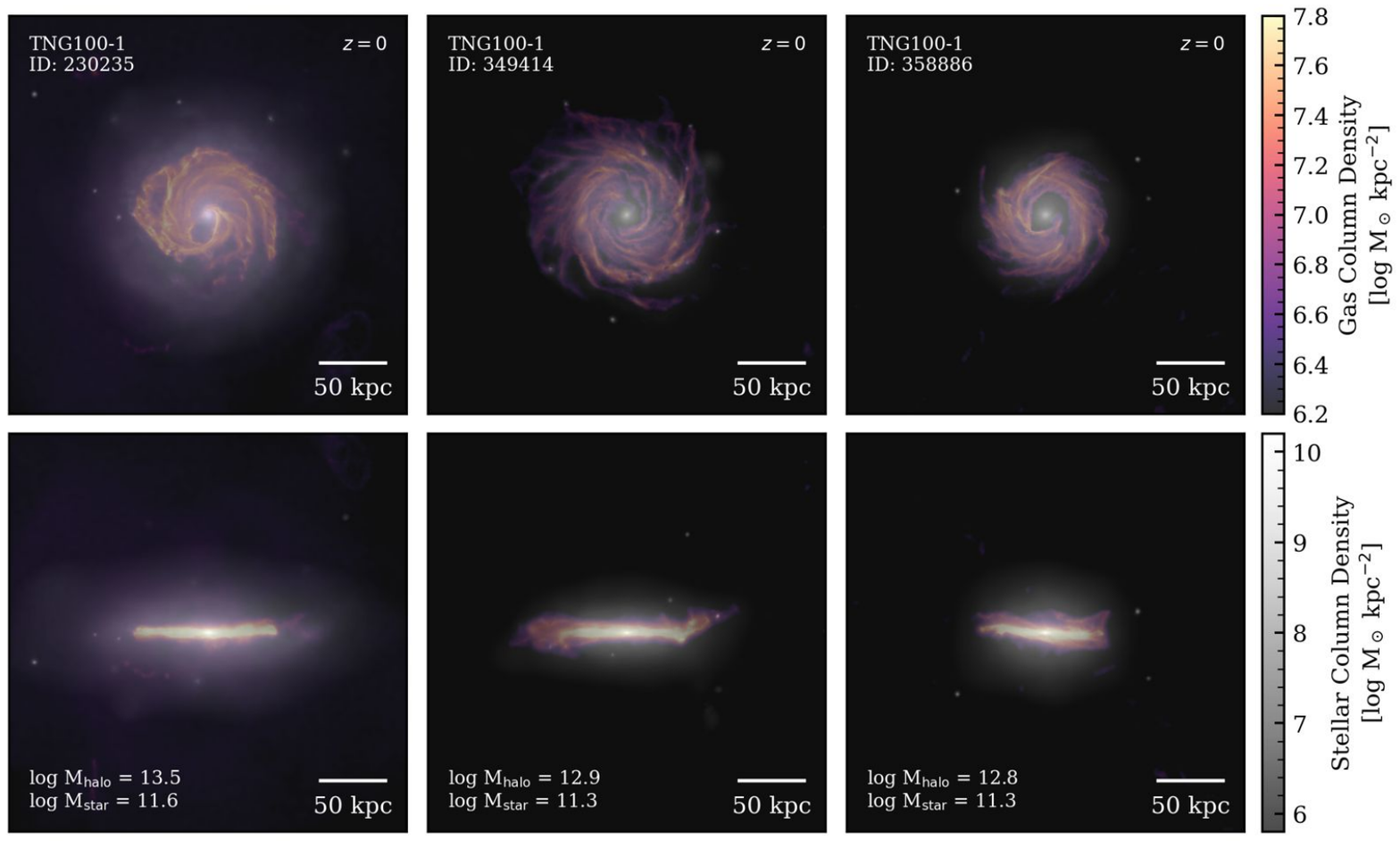} 
\caption{Blended stellar and gas density of three massive disks in face-on (top row) and edge-on (bottom-row) views, selected using a $\lambda/\sqrt{\varepsilon} >$ 0.71 threshold. All galaxies have a stellar mass log$_{10}M_\star \gtrsim 11.3$[M$_\odot$].}

\label{fig:example_disks}
\end{figure*}

\subsection{Kinematical properties}
\label{sec:kinematics}

This subsection explains the methodology employed to measure the kinematic properties for categorizing galaxies into disks and spheroids. We followed the procedure proposed in \citet[][]{Lagos17} to measure, from the stellar component, the velocity dispersion ($\sigma$); the specific angular momentum ($j$); the rotational velocity ($V_{rot}$); and the spin parameter $\lambda_{R}$, defined below. 
Unless stated otherwise, all properties were estimated within three stellar half-mass radii, i.e., three times the radii enclosing half of the galaxy's stellar mass. This property is used as a proxy for the 'half-light radius' of the galaxy.

Firstly, we computed the specific angular momentum of the stellar component, $j_{\star}$, using the formula:

\begin{equation}
    j_{\star} = \frac{\sum_i m_i (r_i - r_{\rm{COM}})\times (v_i - v_{\rm{COM}})}{\sum m_i}
\end{equation}

Here, $r_i$, $v_i$ and $m_i$ represent the position, velocity and mass of the $i$-th particle of a galaxy, while $r_{\rm{COM}}$ and $v_{\rm{COM}}$ are the position and velocity of the center of mass of the galaxy, respectively.
With the specific angular momentum, the rotational velocity of a galaxy, $V_{\rm rot}(r)$, can be determined as:

\begin{equation}
    V_{\rm rot}(r) \equiv \frac{\left| j_\star(r)\right|}{r} 
\end{equation}

Additionally, we can measure the velocity dispersion perpendicular to the disk's mid-plane. To do this, we will take the velocity dispersion above parallel to the total stellar angular momentum ($L_{\star})$ from the galaxy. In the measurement of $L_{\star}$, we consider all stellar particles bound to the subhalo. 

Finally, the 1D velocity dispersion parallel to the disk is measured as follows: 

\begin{equation}
    \sigma_{1,D} = \sqrt{\frac{\sum_i m_i (\Delta v_i cos\theta_i)^2 }{m_i}}
\end{equation}

Here, $\Delta v_i = \left| v_i - v_{\rm{COM}} \right|$ is the velocity relative to the centre of mass, and $\cos\theta = \frac{\Delta v_i \cdot L_{\star}}{\left| v_i\right|\left| L_{\star}\right|}$.

The $\lambda_R$ parameter was presented by \citet{Emsellem07} to estimate how rotationally supported a galaxy is.
As shown in \citet{Naab14}, to measure $\lambda_R$ appropriately, we need to combine the properties mentioned above in annuli as follows:

\begin{equation}
    \lambda_R = \frac{\sum_{i=1}^{N(r)} m_{\star,i}r_i V_{rot}(r_i)}{\sum_{i=1}^{N(r)}\sqrt{V_{rot}^{2}(r_i) + \sigma_{1D,\star}^{2}(r_i)}}
\end{equation}

Here, the sum is over all radial bins, N(r), from the inner one to r, and $m_{\star, i}$ is the stellar mass enclosed in each bin.
Although it is evident that, by design, this property is dependent on the resolution of the bins, our focus on studying kinematical properties in the internal parts of the galaxies in our sample ensures that our results remain consistent when using a fixed aperture of three half-mass radii for all the properties mentioned above, including the $\lambda_R$ parameter.

Additionally, we gauge the galaxy's ellipticity within the same enclosed region using only the stellar component's inertia tensor.
The components of the inertia tensor are measured as follows:

\begin{equation}
\begin{split}
    I(r)_{xx} = \sum_{i} m_{i}(y^{2} + z^{2}) \\
    I(r)_{xy} = -\sum_ix_iy_i
\end{split}
\end{equation}

Due to this tensor's symmetry, measuring the other components follows the same procedure. After constructing the tensor, it is diagonalized, and the eigenvalues are sorted in ascending order. These eigenvalues are then utilized to define the three axes of the galaxy.

\begin{equation}
\begin{split}
    a &= \sqrt{\lambda_2+\lambda_3-\lambda_1} \\
    b &= \sqrt{\lambda_1+\lambda_3-\lambda_2} \\
    c &= \sqrt{\lambda_1+\lambda_2-\lambda_3}
\end{split}
\end{equation}

Finally, we define the ellipticity of the galaxy, $\varepsilon$, as:
\begin{equation}
    \varepsilon = 1 - \frac{c}{a}
\end{equation}

Where a and c correspond to the major and minor axis, respectively.
-

\section{Results}
\label{sec:results}

Massive disks are rare in the Universe, with previous studies reporting that superluminous spiral galaxies, a type of massive disk, account for no more than $\sim 10\%$ of galaxies at the high-mass end (log$_{10}M\star \geq 11.4$[M$_\odot$]) \citep[][]{Jackson20, Jackson22}. Additionally, \citet[][]{Quilley22} suggests an upper threshold in stellar mass for disks to exist. Notably, their work found no disks more massive than log$_{10}M\star \geq 11.8$[M$_\odot$]. The reasons behind this apparent threshold in stellar mass remain an open question.
To address this, in the first part of this article, we will select all disks within the simulation using the morpho-kinematic statistics detailed in Section \ref{sec:kinematics}. We choose galaxies dominated by their rotational support and exhibiting small ellipticities by applying a threshold in the ratio between $\lambda$ and $\varepsilon$.
Figure \ref{fig:disk_fraction} illustrates the disk fraction as a function of stellar mass for all galaxies in the simulation. We used three different threshold criteria ($\lambda / \sqrt{\varepsilon} = 0.31; 0.71; 0.81$ represented in red, green, and blue bars, respectively) to select disks. These thresholds correspond to the suggested values in \citet[][]{Emsellem11} for different apertures. For simplicity, we refer to fast and slow rotators as disk and spheroidal galaxies, respectively, as those galaxies above (below) our kinematic selection threshold.
It is important to note that for our final disk classification, we verified the morphologies of the selected disks through visual inspection to ensure that our morpho-kinematic criteria identified actual disks.

From Figure \ref{fig:disk_fraction}, it is apparent that the fraction of disks decreases with increasing mass. The disk fraction steadily decreases at log$_{10}M\star \geq 10.5$[M$\odot$]. It is worth noting that irrespective of the selected threshold, there are disks with stellar masses up to log$_{10}M_\star \geq 11$[M$\odot$], with the most extreme case reaching log$_{10}M_\star \sim 12$[M$_\odot$].

Figure \ref{fig:example_disks} shows the gas density of some of the most spectacular cases
in our sample, all with log$_{10}$M$_\star \geq$ 11.4[M$_\odot$]. 
This mix of spirals and lenticular galaxies resides in various environments, from the field to clusters and satellites to centrals. 

For the remainder of this article, we will refer to as ``Supermassive Disks'' all disks with stellar masses log$_{10}M_\star \geq 11$[M$_\odot$], where the drop in number of disks in IllustrisTNG is noticeable.
Also, from the three aforementioned thresholds, we will focus only on $\lambda / \sqrt{\varepsilon} = 0.31$ and 0.71, since there are no substantial differences between using
$\lambda / \sqrt{\varepsilon} = 0.71$ and 0.81, only increasing statistical noise. 
Of the 694 galaxies available within this mass range, we recovered a total of 220 ($32\%$) and 50 ($7\%$) disks for $\lambda / \sqrt{\varepsilon} = 0.31$ and 0.71, respectively.

\subsection{Control Sample}
Constructing a control sample is crucial in this project. It plays a pivotal role in isolating the phenomena that lead to the formation of supermassive disks, thereby enhancing the accuracy and reliability of our findings. We can identify intrinsic properties and behaviors specific to our galaxies of interest by comparing massive fast and slow rotators. Different methods of building a control sample can introduce selection biases, leading to misleading interpretations of the data. To mitigate these issues, it is essential to impose constraints on parameters such as redshift, stellar masses, and local densities, which can significantly reduce selection biases.
To do this, we build a control sample of massive spheroidal galaxies at $z=0$, with the same stellar mass distribution as our supermassive disks, following \citet{Perez09}, i.e.:  
\begin{itemize}
    \item Select a candidate sample of spheroidal galaxies with log$_{10}$M$_{\star} \geq$11 and $\lambda / \sqrt{\varepsilon} < 0.1$.
    \item For each massive disk, look for massive spheroids with the same stellar mass with 0.01dex of tolerance (abs(log$_{10}$M$_{\star, disk}$ - log$_{10}$M$_{\star, Sph}$ < 0.01)) to construct a sample of candidates.
    \item Randomly select galaxy analogues per supermassive disk to build our control sample while maintaining the stellar mass distribution.
\end{itemize}
By constructing our control sample, we ensure that we compare the same stellar mass distribution while maximizing the number of candidates to compare. In this case, depending on the morpho-kinematic threshold imposed for the supermassive disks, we end up with 220 and 250 massive spheroids in the control sample for disks with $\lambda / \sqrt{\varepsilon} = 0.31$ and  0.71, respectively.
\begin{figure}
\centering
\includegraphics[width=0.5\textwidth]{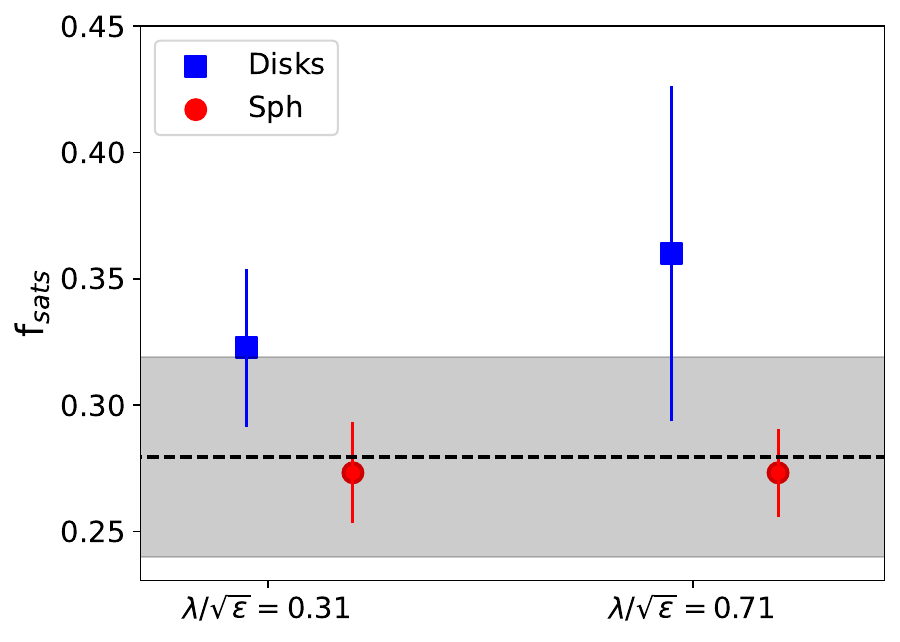}
\caption{Fraction of satellites in our sample of supermassive galaxies. Disks and spheroids are shown in squares (blue) and filled circles (red). The black dashed line shows the mean fraction of satellites within the massive regime. Errors correspond to one sigma binomial errors as stated in \citep{Cameron11}, while the gray shaded area corresponds to three sigma binomial errors for the whole sample. We can see that although disks tend to have a mild tendency to be satellites, the fractions of satellites for disks and spheroids are comparable within one sigma.}

\label{fig:sat_frac_disksph}
\end{figure}

\begin{figure*}
\centering
\includegraphics[width=\textwidth]{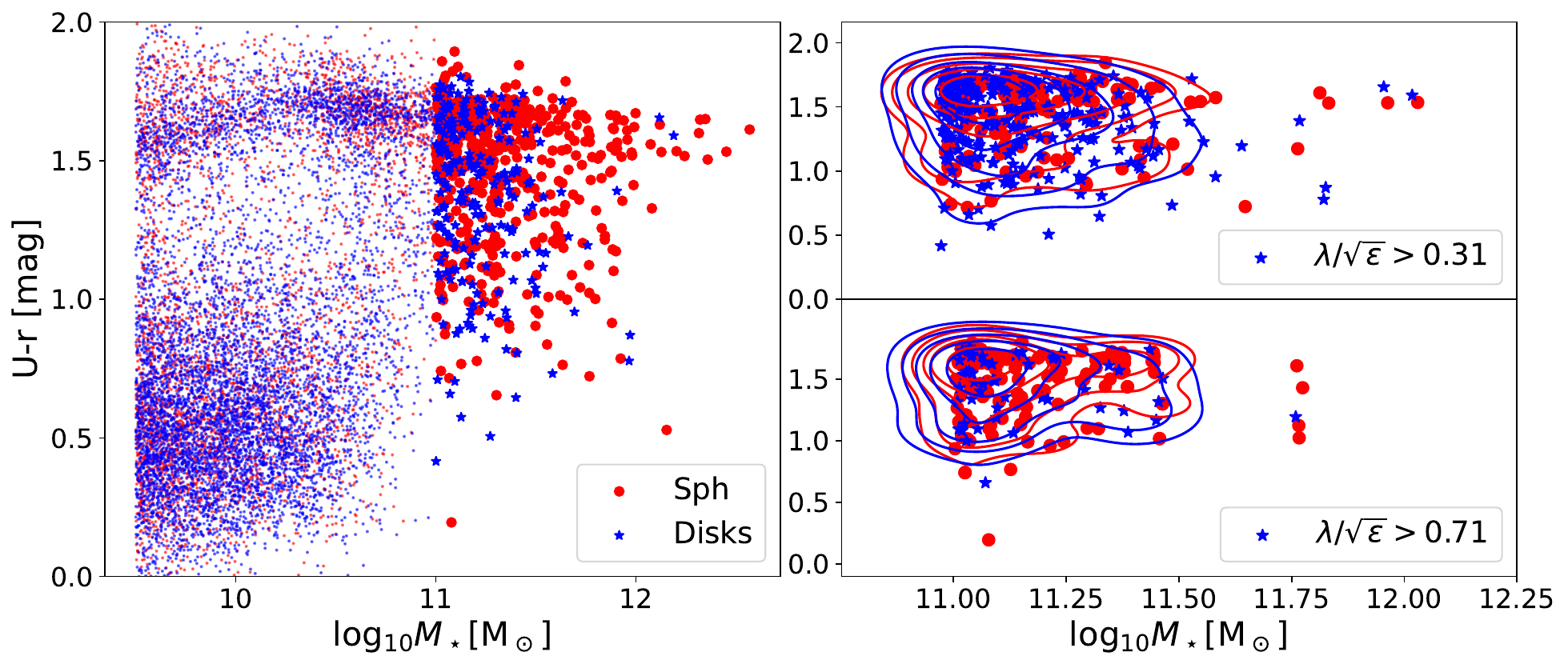}
\caption{Left: Colour (U-r) stellar mass diagram for all galaxies in the simulation. Blue and red dots show disk and spheroidal galaxies, with stars highlighting galaxies with log$_{10}M_\star \geq 11$. On this panel, we choose a threshold $\lambda/\sqrt{\epsilon} \geq 0.31$ to split between disks and spheroids. 
Right: Selection of massive galaxies in our sample. The upper and lower panels use two different selection criteria: $\lambda/\sqrt{\epsilon} \geq 0.31$ and 0.71, respectively. Stars (blue) and dots (red) correspond to disks and control samples associated with each sample, respectively. Contours are added for visualization purposes.
Disk galaxies show a broad distribution in mass and color, but their density decreases dramatically toward larger stellar masses as expected, populating the green valley preferentially for log$_{10}M_\star \geq 11$[M$_\odot$]. Spheroidal galaxies dominate the red population at any stellar mass but can also be found as blue galaxies at low stellar masses.}
\label{fig:colour_mass}
\end{figure*}

\begin{figure}
\includegraphics[width=0.5\textwidth]{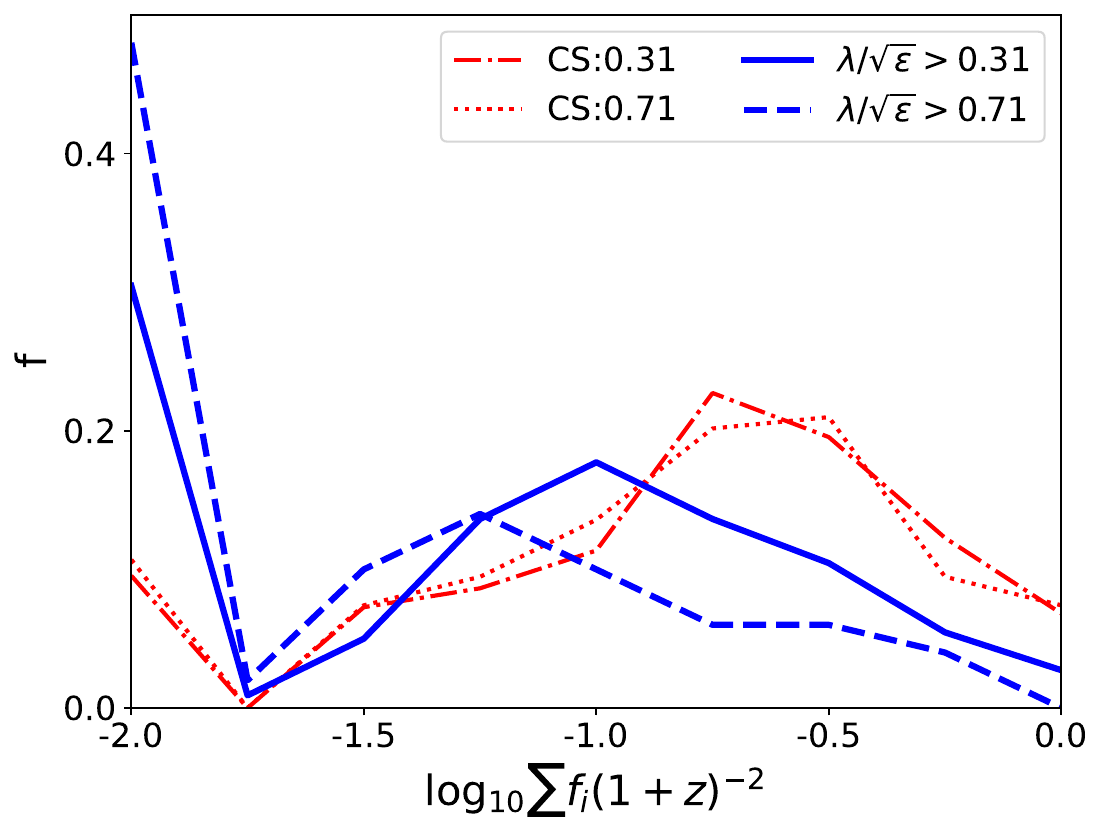}
\caption{Relative merger importance for the growth in stellar mass of massive galaxies. Continuous (dash-dot) and dashed (dotted) blue (red) lines correspond to disks (spheroids; control sample) selected with $\lambda/\sqrt{\varepsilon} > 0.31$ and 0.71, respectively. 
Disks exhibit a quiescent merging history, indicating that most of their stellar mass was formed through star formation rather than the accumulation of mergers. As we move toward stricter thresholds, the importance of mergers reduces significantly. On the other hand, spheroids require a significant fraction of mergers to form, regardless of the selection criteria used.}
    \label{fig:merger_smass}
\end{figure}

Given that different kinematic criteria produce different stellar mass distributions, we built a control sample for each galaxy set. 
It is worth noting that this random procedure was performed for several control samples, and our results remained unchanged when we selected other galaxy sets. Throughout this article, we present our results based on the same control sample (CS) for each chosen kinematic criterion.
In Figure \ref{fig:sat_frac_disksph}, we show the fraction of satellites for supermassive galaxies split by morphological type using a kinematic threshold of $\lambda / \sqrt{\varepsilon} = 0.31$ and 0.71 to split between disks and spheroids, shown as squares (blue) and filled circles (red). The error bars correspond to binomial errors within one sigma. The black dashed line shows the median satellite fraction for supermassive galaxies, and the grey-shaded area shows errors within three sigma for the whole sample of supermassive galaxies. At first glance, it appears that supermassive disks tend to exhibit a mild tendency to be satellites compared to both spheroids and the entire sample. Nevertheless, we cannot detect any special trend within one standard deviation. We will further explore these results in Section \ref{sec:environment}.

\subsection{Colour-mass diagram}

To visualize the distribution of disks compared to the entire population of galaxies in the simulation, Figure \ref{fig:colour_mass} shows the color-stellar mass relation for all galaxies with log$_{10}M_\star \geq 9.5$ [M$_\odot$] (left panel). The blue and red dots correspond to disk and spheroidal galaxies, respectively. Additionally, we highlighted supermassive galaxies with prominent stars to better visualize their position in the diagram. 
Additionally, in the rightmost panels, we can see a zoomed region including only supermassive galaxies, where disks are selected using $\lambda / \sqrt{\varepsilon} = 0.31$ (upper panel) and 0.71 (lower panel).
As we can see, supermassive disks occupy all regions in the color-stellar mass diagram, mainly populating the green valley and the red population. Although this may seem counterintuitive at first glance, as previously mentioned, our supermassive disks correspond to a mix of spirals and lenticular galaxies that should dominate at this mass range. At the same time, although our galaxies have supermassive disks, most also have a well-defined supermassive spheroidal (red) component within their bulge.

It is worth noting also that although we recover the drop in disk fraction shown in \citet{Quilley22}, it is shifted to lower masses, being at log$_{10}M_\star = 11$[M$_\odot$] and log$_{10}M_\star = 11.4 $[M$_\odot$] for TNG and \citet{Quilley22} respectively. 
This is somewhat expected given that, as shown in \citet{Genel18}, late-type galaxies within IllustrisTNG are slightly less massive than what is expected for a given size at log$_{10}$M$_{\star}/M{_\odot} > 10.5$. Additionally, as reported in our work, there are few late-type galaxies (or fast-rotators) with log$_{10}$M$_{\star}/$M${_\odot} > 11.4$. This result holds for early-type galaxies as well, although there are several early-type galaxies with log$_{10}$M$_{\star}/$M${_\odot} > 11.4$.

\begin{figure*}
\includegraphics[width=\textwidth]{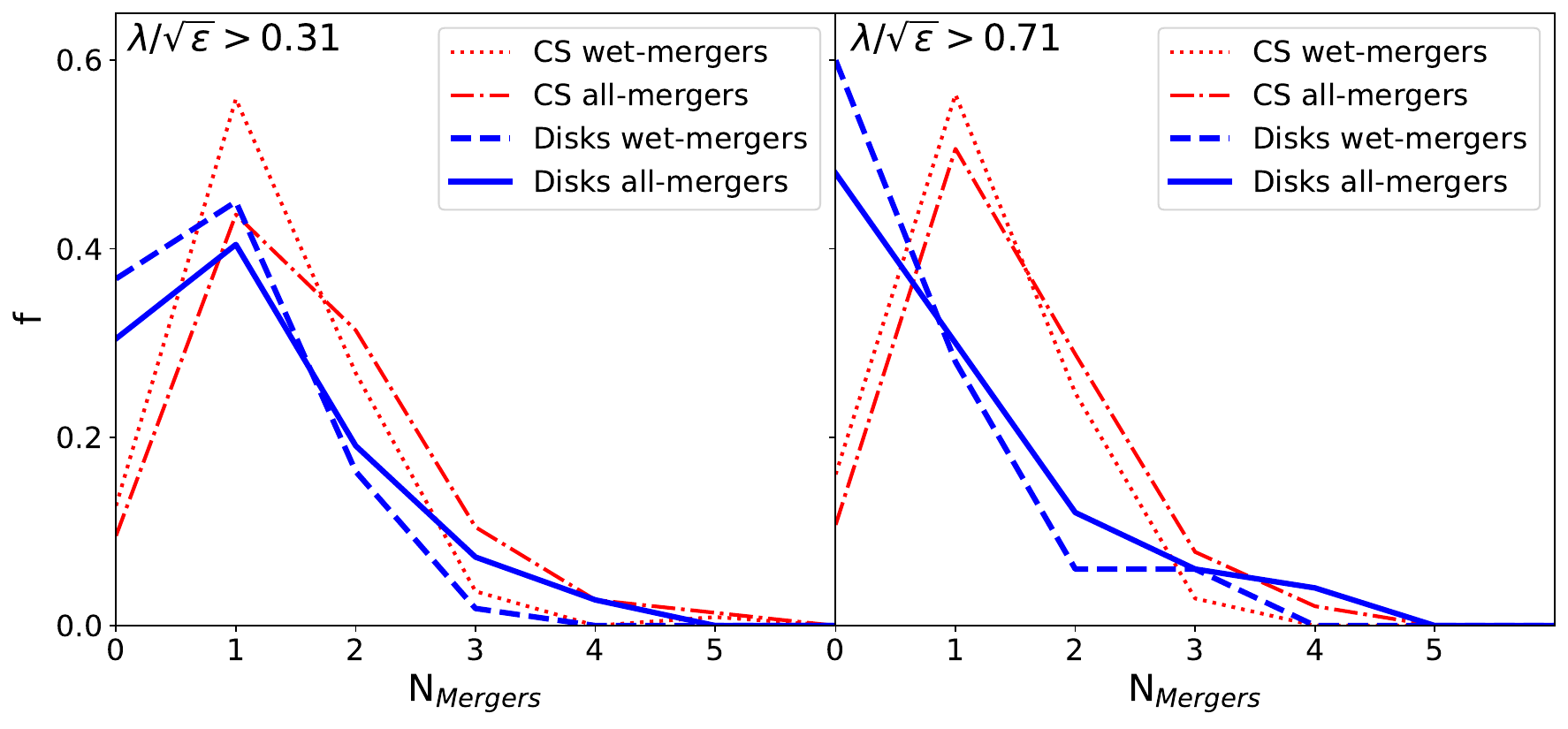}
\caption{Number of mergers experienced by supermassive galaxies. 
Continuous and dashed (blue) lines show the number of all and only gas-rich mergers experienced by disks. In contrast, dash-dot and dotted (red) lines correspond to all and only gas-rich mergers experienced by our control sample, for  $\lambda / \sqrt{\varepsilon} = 0.31$ (left panel) and 0.71 (right panel), respectively. 
Up to 45$\%$ of disks in our sample did not experience any merger at $z<1$, with less than 20$\%$ of disks in our sample experiencing two or more mergers. 
On the other hand, spheroidal galaxies have a very active merging history, with more than 70$\%$ of them experiencing one or two mergers at $z<1$.}
\label{fig:N_mergers}
\end{figure*}

\subsection{Merger histories}
\label{sec:merger}

As a proxy to quantify how essential mergers were for the assembly of our sample, we define a parameter that relates the number of mergers, the stellar mass ratio of galaxies at the time of the merger, and the redshift at which this merger took place, following \citet[][]{Fragkoudi20}. 
Here, we measure the sum of the relative importance of each merger in the build-up of the stellar mass of that galaxy using the following equation: 

\begin{equation}
	\sum \frac{f_i}{(1+z_i)^2}
\end{equation}

where $f_i$ is the ratio between the maximum stellar mass reached by the two merging galaxies; $f_i = M_{\rm sat} / M_{\rm cen}$; and $z_i$ corresponds to the redshift of the merging time. We define the merging time as the last moment when both galaxies can be recognized independently by \textsc{subfind}.
This is done to ensure the merger is final and to avoid cases where \textsc{subfind} could detect two galaxies passing close to each other as a single merged galaxy.
This parameter was measured for all mergers since $z=1$ with $f_i > 1/10$. Then, we sum the parameter obtained at all $z$ for each galaxy. 

Figure \ref{fig:merger_smass} shows the relative importance of mergers for each galaxy for all our thresholds. Continuous and dashed (blue) lines show the merger impact for disks selected with $\lambda/\sqrt{\varepsilon} > 0.31$ and 0.71, respectively. Dash-dot and dotted (red) lines show the merger impact for both control samples. 
For galaxies with no significant mergers at $z<1$, the value of $\sum f_i(1+z_i)^{-2}$ was saturated at $\sum f_i(1+z_i)^{-2} =10^{-2}$ for visualisation purposes.
As shown in Figure \ref{fig:merger_smass}, disk galaxies have a significantly more quiescent recent ($z<1$) merger history than spheroidal galaxies. 
For any selection threshold, more than $30\%$ of disk galaxies did not experience any significant merger at $z < 1$. Moreover, we can see that for $\lambda/\sqrt{\varepsilon} =  0.71$, $46\%$ of galaxies did not experience any substantial mergers since $z=1$. 
Additionally, when we examine each distribution, we can see that the merger's impact on disks is an order of magnitude less significant for supermassive disks than in our control sample. Additionally, the impact of mergers becomes less significant when stricter criteria are considered.
This result highlights that spheroidal galaxies tend to have a more active recent merging history, with a broader range of values for the merging parameter. Moreover, we can see that only $\sim 10\%$ of the galaxies from our control sample did not have any significant merger since $z=1$ at any morpho-kinematic threshold.

To further highlight the differences between populations, Figure \ref{fig:N_mergers} shows the number of significant mergers ($f_i \geq 1:10$) experienced by the galaxies in our sample. In this case, we not only show normal mergers, but we highlight gas-rich mergers as well, where we define a gas-rich merger as mergers where the maximum gas mass of the two merging galaxies, at their merging time, i.e., $f_{gas} = M_{gas, sat}/M_{gas, cen} > 0.1$.
Continuous and dashed (blue) lines show the number of all and only gas-rich mergers experienced by disks. In contrast, dash-dot and dotted (red) lines correspond to all and only gas-rich mergers experienced by our control sample, for  $\lambda / \sqrt{\varepsilon} = 0.31$ (left panel) and 0.71 (right panel), respectively.
Regardless of the threshold used as selection criteria, disks have
a lower number of mergers compared to spheroids.
Moreover, at most, $80\%$ of disk galaxies suffered only one significant merger throughout their history. On the other hand,  $\geq50\% $ of spheroids need at least two mergers to form.

When comparing the number of gas-rich mergers with the whole merging history, we can see that both distributions are very similar. This means that, for disk galaxies, in addition to having fewer mergers than our control sample, the few mergers they encounter are gas-rich.
Moreover, in the case of galaxies with $\lambda / \sqrt{\varepsilon} \geq 0.71$, by comparing cases one by one, we found that only five galaxies experience mergers that do not classify as gas-rich. These cases were minor mergers at $z \sim 1 $, although significant.

These results suggest that gas-rich mergers may allow disks to be destroyed and reformed as supermassive disks. This may not be the case for our control sample, as having a large number of mergers could dynamically heat the stellar population, inhibiting the reformation of disks.
Nevertheless, we want to highlight that even though this may be the case for some galaxies, the bulk of our sample comprises supermassive disk galaxies that do not experience any merger at $z<1$ regardless of their mass.

\subsection{The impact of AGN in forming supermassive disks.}

\begin{figure}
\centering
\includegraphics[width=0.5\textwidth]{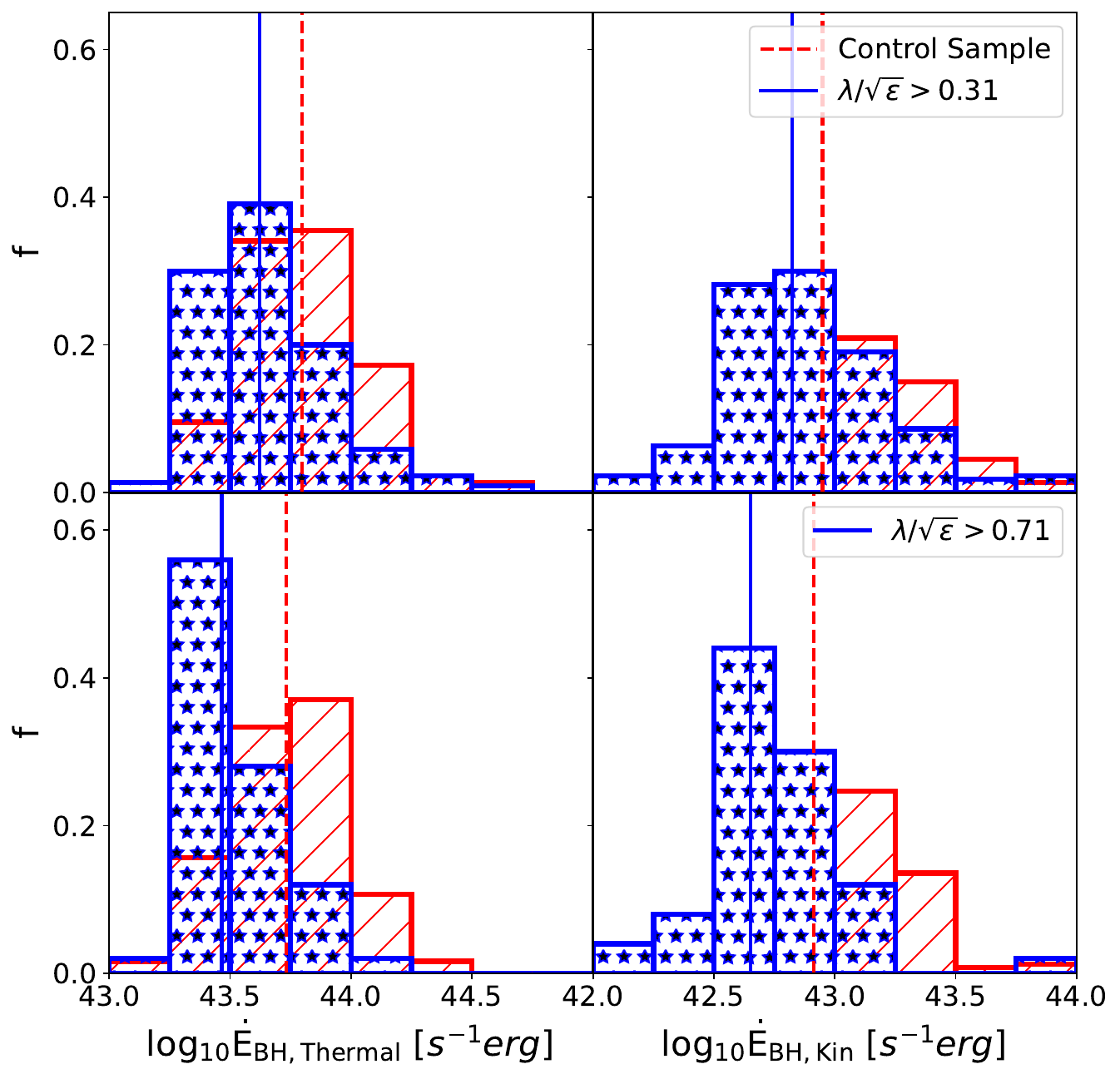}
\caption{Distribution of energy injected into the medium by AGN feedback thermally (quasar-mode, left panels) and kinetically (radio-mode, right panels) for disks (starred blue bars) and our control sample (hashed red bars). Disk galaxies show marginally smaller values of both types of energetic feedback ($\sim 0.125$ - $\sim 0.25$ dex) depending on the morpho-kinematic threshold.}

\label{fig:agn_smass}
\end{figure}

Another possibility to stop disks from growing is if enough energy is injected into the interstellar medium. This injection may prevent disks' reformation and halt their star formation, making them unable to survive in time.
Several feedback mechanisms may inject energy into the medium, but the most important one, especially at the high-mass bin, is the AGN feedback, as extensively reviewed in \citet{Fabian12}. To see the impact that the AGN feedback may have on the survival conditions of supermassive disks, we track the energy injected into the medium thermally and kinetically in those phases where the AGN was active as a quasar and in radio mode, respectively. 
Figure \ref{fig:agn_smass} shows the AGN's mean energy injected into the galaxy in quasar (thermal, left panels) and radio (kinetic, right panels) modes for our disks (starred blue bars) and our CS (red hashed bars). The energy feedback distribution for each of our selected thresholds is shown from top to bottom.
The continuous (blue) and dashed (red) lines correspond to the median for each distribution for our supermassive disks and CS.
We can see that the differences between samples are small for both feedback methods, kinetic and thermal AGN, with supermassive disks having slightly less active AGN histories ($\sim 0.25$ dex at $\lambda / \sqrt{\varepsilon} > 0.71$).
Although a small difference related to the morphology may be seen in the thermal energy injected throughout galaxy history, it is worth mentioning that the kinematic mode is expected to be the one that directly affects the morphology of galaxies, while the thermal mode at most may affect their star formation history, as shown in previous work \citep[eg.][]{Pillepich18a, Rosas-Guevara19}. In this sense, the thermal mode may stall the star formation in disks, transforming them from spirals to lenticulars, but it is not expected to transform a disk into a spheroid. Considering this, we conclude that AGN does not play a key role in stopping the formation of supermassive disks but may stall their growth in stellar mass or prevent the reformation of a new disk after a significant merger.

\begin{figure}
\centering
\includegraphics[width=0.5\textwidth]{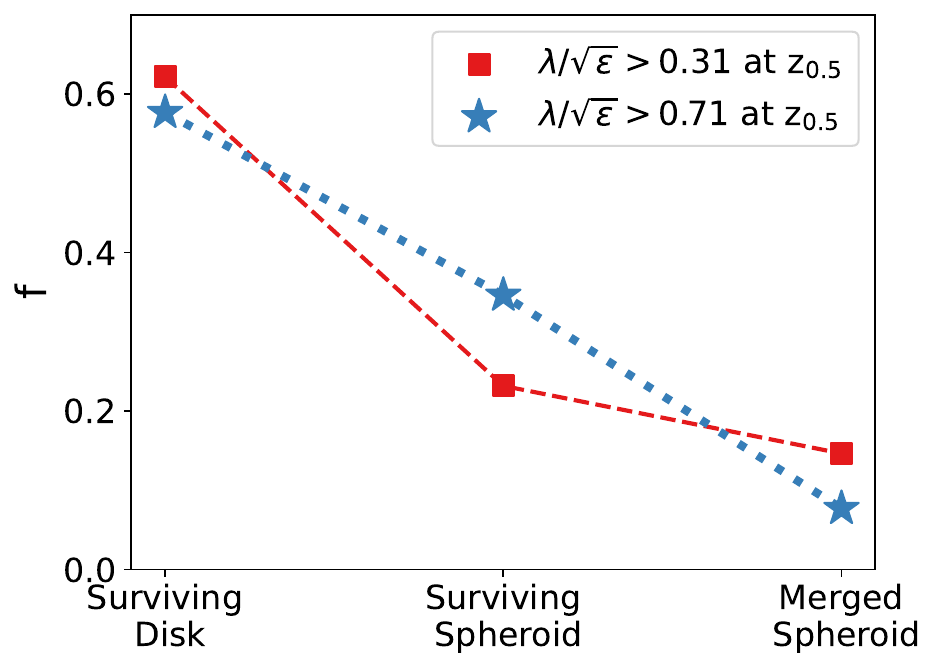}
\caption{Fate of supermassive disks (log$_{10}M_\star ~ 10^{11}$[M$_\odot$]) selected at $z=0.5$. The fraction of surviving disks (SD), the fraction of disks that merge with a more massive galaxy (MS), and the fraction of disks that transform to spheroidal throughout their history (SS) are shown here. We show disk selected by different thresholds in red and blue lines with $\lambda / \sqrt{\varepsilon}$ = 0.31 and 0.71, respectively. }

\label{fig:disk_z05}
\end{figure}

\begin{figure*}
\centering
\includegraphics[width=\textwidth]{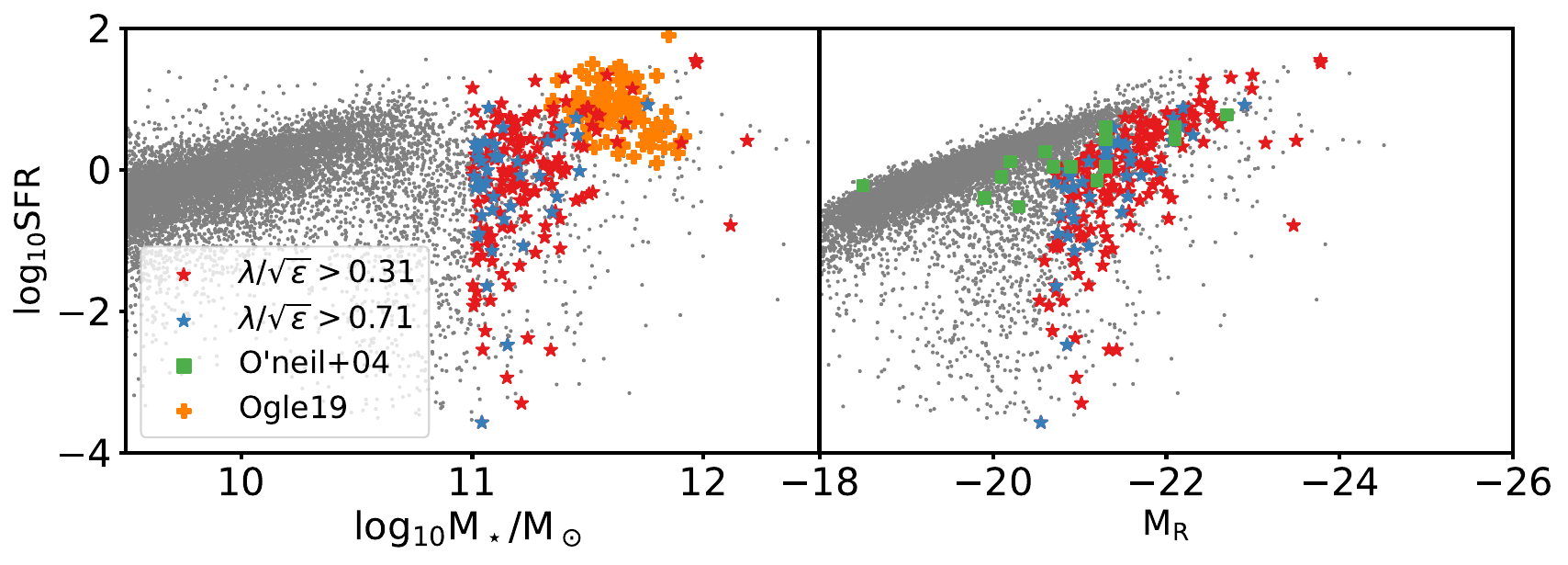}
\caption{Star formation rate (SFR) as a function of stellar mass (left) and absolute magnitude (right) for our simulated supermassive disk galaxies compared with observational samples. In the left panel, our galaxies are compared with the supermassive galaxy sample from \citet{Ogle19} (orange crosses). In contrast, the right panel shows a comparison with the low surface brightness (LSB) galaxies from \citet{Oneil04}. Most observational galaxies lie along the star-forming main sequence, whereas our simulated sample includes both star-forming and quiescent systems, highlighting their broader diversity in star formation activity.}

\label{fig:Obs_comp}
\end{figure*}

\begin{figure}
\centering
\includegraphics[width=0.5\textwidth]{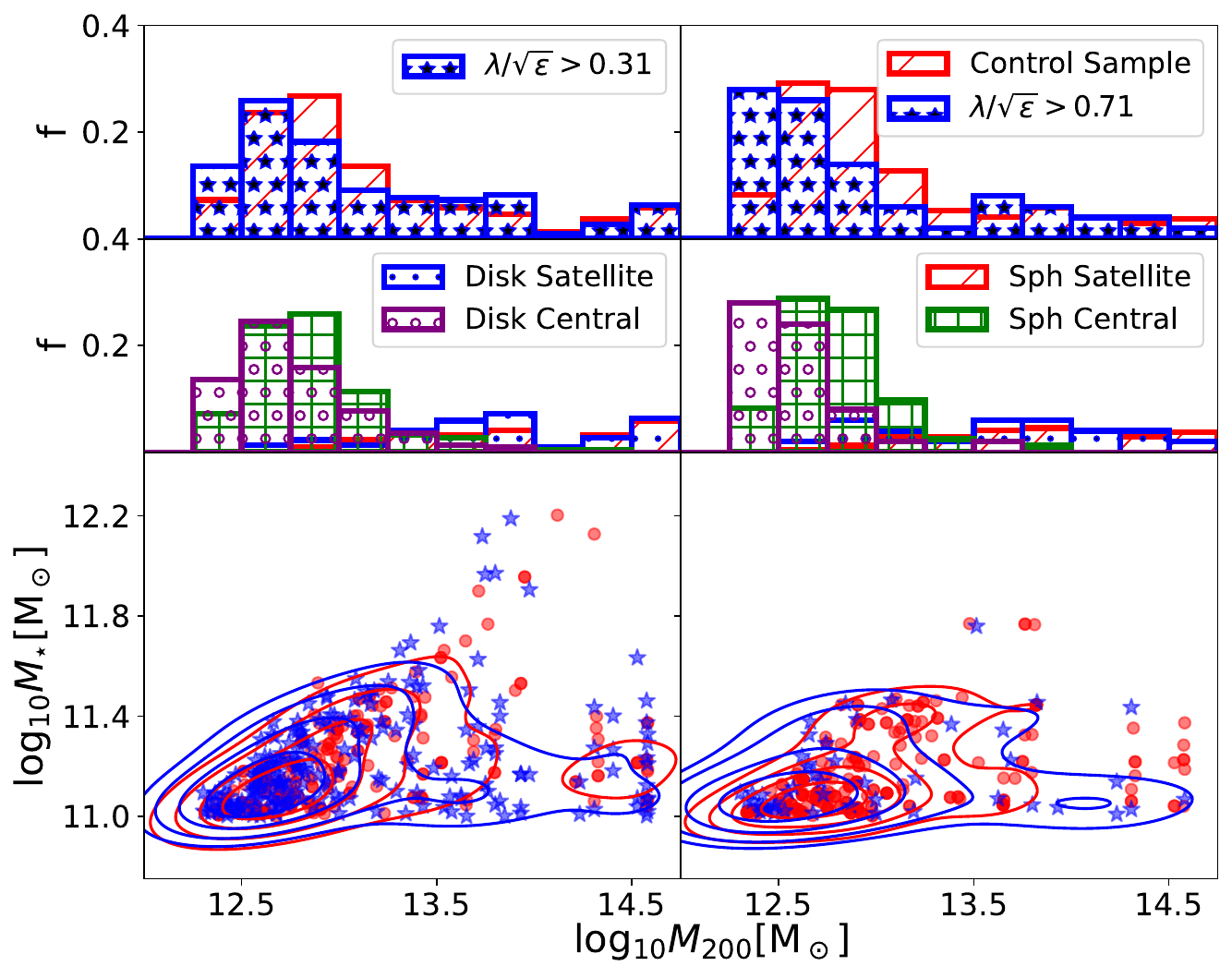}
\caption{Distribution of haloes versus stellar mass for the galaxies in our sample, using a threshold of $\lambda/\sqrt{\epsilon} \geq 0.31$ (left) and 0.71 (right).
Stars (blue) and dots (red) represent disks and spheroids (control sample), respectively. In the upper panels, we plot a histogram showing the log$_{10}$M$_{200}$ mass distribution in starred (blue) and dashed (red) bars for disks and spheroids, respectively. Additionally, we show the distribution of central and satellite disks in open dots (blue) and circles (purple) and central and satellite spheroids in squared (green) and lined (red) histograms.
We can see that disk galaxies do not have any particular distribution in M$_{200}$, showing even a slightly extended distribution in halo mass than our control sample, reaching up to the most massive clusters. }

\label{fig:m200_dist}
\end{figure}

\begin{figure}
\centering
\includegraphics[width=0.5\textwidth]{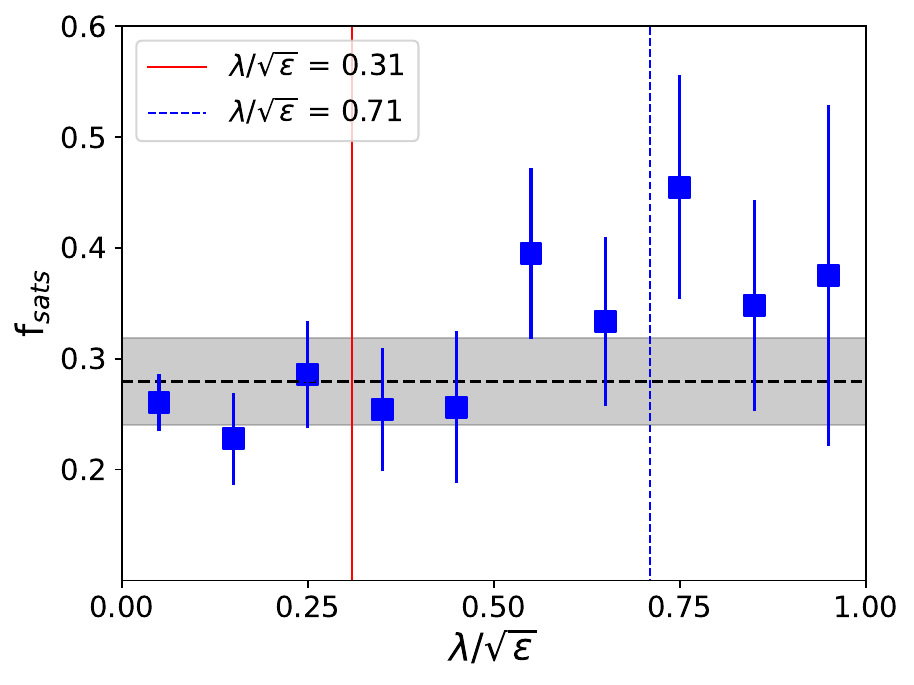}
\caption{Fraction of satellites in our sample of supermassive disks, compared to the whole sample of massive spheroids. We can see that the fraction of supermassive disks as satellites is similar to their spheroidal counterpart.}

\label{fig:sat_frac_lambda}
\end{figure}
\section{Discussion}
\label{sec:discussion}

\subsection{Evolution of supermassive disks in the last 4Gyr}
\label{sec:evolution_disks}
Supermassive disks have experienced a more quiescent merging history than spheroidal galaxies. However, it is still unclear how likely they are to survive once they have formed.
To answer this question, we made a new sample of supermassive disks selected at $z=0.5$ and followed their evolutionary history up to $z=0$. 
For these new disks, we use the same constraints as used for our disks at $z=0$, i.e., log$_{10}M_\star \geq 11$[M$_\odot$] and $\lambda / \varepsilon \geq 0.31$ and 0.71, with a total of 164 and 26 supermassive disks, respectively at $z=0.5$.

Surprisingly, our results suggest that most disks survive until $z=0$, retaining their disk condition.
Figure \ref{fig:disk_z05} shows the fraction of supermassive disks that survive since $z=0.5$ (Surviving Disks; SD),  the fraction of galaxies that survive until $z=0$, but their morphology was transformed from disk to spheroidal (Surviving Spheroids; SS) and the fraction of disks that do not survived until $z=0$, because they merged with other more massive spheroidal galaxy between $z=0.5$ and $z=0$ (Merged Spheroid; MS),. 
When selecting disks with $\lambda/\sqrt{\varepsilon} = 0.31$ and  0.71, we found that: 
\begin{itemize}
    \item 102 (63$\%$) and 15 (58$\%$) disks survived for at least 4Gyr (Surviving Disks; SD). 
    \item 38 (23$\%$) and 9 (35$\%$) transform into spheroids by $z=0$ (Surviving Spheroid; SS).
    \item 24 (14$\%$) and 2 (7$\%$) merged with a more massive galaxy during this time (Merging Spheroid; MS).
\end{itemize}
It is worth noting that although we found very few supermassive disks with large rotational support at $z=0.5$, this is a consequence of the lack of supermassive galaxies at intermediate $ z$ within our limited volume sample, and not the lack of rotationally supported galaxies.
These results suggest that once supermassive disks are formed, they are long-lived, with $\sim 60\%$ of supermassive disks selected at $z=0.5$ surviving until $z=0$ regardless of the selection threshold. Between 20$\%$-35$\%$ of supermassive disks formed at $z=0.5$ merged with a less massive galaxy, transforming its morphology by $z=0$.
Additionally, it is worth noting that in some cases, disks found at $z=0.5$ may undergo a gas-rich merger, resulting in a transitional state with a broken disk. Nonetheless, if the mergers are gas-rich enough, the disk can be reconstructed, and the morphology may be preserved. This phenomenon will be explored in Section \ref{sec:formation}.

\begin{figure}
\centering
\includegraphics[width=0.5\textwidth]{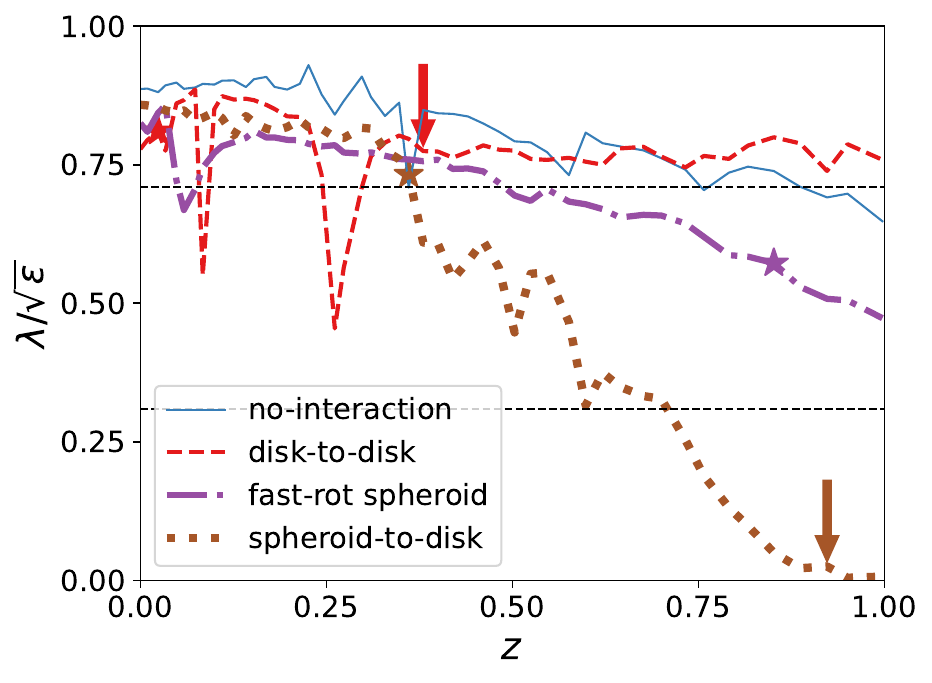}
\caption{Examples of galaxies experiencing different formation pathways: A quiescent merging history (cyan), a disk that experienced mergers (red), a spheroidal galaxy with high rotational support (purple), and a spheroidal galaxy that rapidly transformed into disks after a gas-rich merger (brown). Colored arrows and stars highlight the time when the merger starts and ends.}

\label{fig:path_examples}
\end{figure}

\subsection{A comparison with observations}
\label{sec:obs}
Supermassive disks are rarely found in the Universe, with $\leq 200$ objects reported until now \citep{Ogle19, Jackson22}. As we show in Section \ref{sec:results}, they only constitute between  $\sim 7 - 30 \%$ of the galaxies in the TNG sample, depending on the criterion. Moreover, in \citet[][]{Jackson20}, using a different set of simulations, they found that supermassive disks constitute $\sim 11\%$ of their sample. While this value exceeds what we find under our more strict selection criteria, it still corresponds to only a minority of the massive systems.
Additionally, in \citet[][]{Jackson22}, it is reported that supermassive disks account for $\sim 13\%$ of galaxies within the high-mass end, in agreement with their previous work.
On the other hand, \citet{Zhu23} studied a particular type of supermassive disks, the Giant Low Surface Brightness galaxies (GLSB), using TNG-100. They found that around $6\%$ of galaxies per stellar mass bin, within the mass range of 10.2 $\leq$log$_{10}$M$_\star/$M$_\odot\leq 11.6$, correspond to GSLB. This is in rough agreement with our results, although the studied mass range differs. 
Additionally, in \citet{Perez-Montano19}, using a galaxy sample from the KIAS Value-Added Catalogue \citep[KIAS VAC, a subsample from Sloan Digital Sky Survey Data Release 7 SDSS DR7][]{Abazajian09, Choi2010}, they showed that low-surface brightness galaxies (LSBG) tend to reside closer to filaments than clusters, compared to high-surface brightness galaxies (HSBG). They further tested these results using IllustrisTNG-100 \citep{Perez-Montano22, Perez-Montano24}, where they found that, in addition to being closer to filaments, rotationally supported LSBGs tend to reside in low-density environments. They suggest that this environment may contribute to the formation of this type of object.

In Figure \ref{fig:Obs_comp}, we compare our sample of simulated supermassive disk galaxies with two observational datasets: the supermassive disks from \citet{Ogle19} (orange crosses, left panel) and the low surface brightness galaxies (LSBGs) from \citet{Oneil04} (green squares, right panel). We show the star formation rate (SFR) as a function of stellar mass (left panel) and absolute R-band magnitude (right panel). Red and blue stars show the sample of supermassive disks selected with a threshold $\lambda/\sqrt{\varepsilon} > 0.31$ and 0.71, respectively.
As shown in Figure \ref{fig:Obs_comp}, most of the observational galaxies lie along the star-forming main sequence. At the same time, our simulated population spans a broader range in SFR, including both star-forming and quiescent systems. This diversity likely reflects the nature of our selection criteria and the fact that our sample is volume-complete. Since we select galaxies solely based on their stellar mass, our analysis is unaffected by observational biases, allowing us to include low-luminosity. These high-mass galaxies may remain undetected in flux-limited surveys.
Although our sample extends to somewhat lower stellar masses than those in \citet{Ogle19}, the star-forming galaxies in our simulation follow a similar SFR–mass relation to that observed. This consistency suggests that the observational sample of supermassive disks may be incomplete, particularly at lower surface brightness. With the advent of next-generation wide-field surveys such as the Vera C. Rubin Observatory's Legacy Survey of Space and Time (LSST), we expect this gap to be significantly reduced, enabling more comprehensive studies of the supermassive disk galaxy population.

\subsection{Environmental distribution}
\label{sec:environment}

Previous studies have shown that these galaxies comprise a minority of the massive galaxies in the Universe. Nevertheless, understanding which environments are more prevalent may help future studies by facilitating their discovery and observation in all-sky surveys.

In Figure \ref{fig:m200_dist}, we show the stellar mass as a function of the host halo $M_{200}$ in which massive galaxies reside. Stars (blue) and dots (red) correspond to our supermassive disks and CS, respectively. The upper histograms correspond to the distribution of the host halo mass in which our galaxies reside, for both thresholds of $\lambda / \varepsilon \geq 0.31$ (left panel) and 0.71 (right panel). 
Additionally, the middle panels display the same halo distribution by separating disks and spheroids into central (purple open circles and green squares) and satellite (open blue dots and red lines) categories, respectively.
Density corresponds to 20$\%$ confidence iso-contours.
Here, we can see that surprisingly supermassive disks inhabit haloes of a wide range of masses. Qualitatively speaking, they tend to inhabit slightly less massive halos than spheroidal galaxies, but exhibit a wider distribution than our control sample, as indicated by the isocontours. Nevertheless, both samples have a broad range of haloes in which they can reside.
Additionally, supermassive disks tend to reside as centrals, following the trends for most supermassive galaxies. It is worth mentioning that satellite supermassive disks tend to reside in massive groups and galaxy clusters. Figure \ref{fig:m200_dist} also highlights that, as we move toward a stricter kinematic threshold, our sample of disks tends to be slightly less massive, but we reckon that this may be related to our limited sample.

Even though the number of supermassive disks decreases with host halo mass, supermassive disk galaxies can reside in any environment, ranging from isolated systems to the most massive clusters in the simulation. It is worth reminding the reader that, as we are defining disks, these massive systems that reside within galaxy clusters as satellites probably correspond to lenticular galaxies and red spirals. As highlighted by Figure \ref{fig:colour_mass}, as we move toward more massive galaxies, disks tend to reside within the green valley and the red population, highlighting that what dominates the formation of rotationally supported galaxies is their merging history instead of their environment.
When splitting the available environments into low-mass groups ($12 \leq$log$_{10}M_{200}$[M$_\odot$]$ \leq 13$), massive groups ($13 \leq$log$_{10}M_{200}$[M$_\odot$]$ \leq 14$) and galaxy clusters ($14 \leq$log$_{10}M_{200}$[M$_\odot$]), we found that around $\sim 60\%$ of supermassive disks reside in low-mass groups, $\sim 25\%$ reside in massive groups, while the remaining $\sim 15\%$ reside in galaxy clusters.

Additionally, their conditions as satellites or centrals may foster or halt their formation and survival. As shown in Figure \ref{fig:sat_frac_disksph}, disks do not have any preferential tendency to be a satellite regardless of the selection criteria. Nonetheless, this was done in general terms. To investigate this relation further, in Figure \ref{fig:sat_frac_lambda}, we show the fraction of satellites as a function of their rotational support. We measured the median fraction of satellites in bins of 0.1$\lambda / \sqrt{\varepsilon}$, shown in blue squares.
The error bars correspond to the binomial errors following \citet{Cameron11} within one sigma. Additionally, the black dashed line shows the average fraction of satellites including all supermassive galaxies, with the gray shaded area showing errors within three sigma. The continuous (red) and dashed (blue) vertical lines highlight the threshold used throughout this work for visualization purposes.
As we move toward higher values of $\lambda / \sqrt{\varepsilon}$, the fraction of satellites grows, with an apparent increase at $\lambda / \sqrt{\varepsilon} > 0.5$. These results at first may suggest that more rotationally supported systems prefer to be satellites instead of central galaxies, in agreement with our previous claim of having a less active merging history. Nevertheless, within one sigma error, we cannot conclude any particular difference with confidence.
It is worth noting that our sample is limited in volume, being a simulation box of $100cMpc$ with only 14 clusters, being the most massive $M_{200} = 3.8\times10^{14}$[M$_\odot$].
Considering this, we acknowledge that the fractions and conditions of centrals and satellites may change if we examine more massive clusters or larger samples of galaxies and environments, where the number of rotationally supported galaxies could increase, reducing the errors.
Nonetheless, our results highlight that the most crucial condition for supermassive disks to exist is a relatively quiescent but gas-rich merger history, which does not correlate directly with the environment.

\subsection{Formation pathways of supermassive disks}
\label{sec:formation}

In \citet[][]{Jackson20}, two different formation scenarios were discussed for the formation of supermassive spirals.
The first one, and the most predominant in their sample, explains that when a massive dispersion-dominated galaxy undergoes a significant merger ($f_{\rm merger} > 1/10$) with a gas-rich satellite, the interaction may spin up the spheroidal component leaving a stellar disk as the remnant of the interaction.
This pathway accounts for $70\%$ of the galaxies in their sample. Additionally, they found that this mechanism is facilitated by galaxies inhabiting less dense environments, which makes them less susceptible to gas-depleting processes like ram pressure and tidal stripping. 
On the other hand, the second formation scenario is due to an anomalously quiescent merging history in their sample, with the remaining $30\%$ retaining their disk condition throughout the galaxy's history.

As we have seen throughout this work, 
although the formation pathways proposed by these authors align with our findings, the fractions that each of these mechanisms accounts for in our sample are different, with most of our supermassive disks experiencing a relatively quiescent merging history compared with spheroidal galaxies. 
It is worth noting that some of our disks have experienced (between 30$\%$ and 40$\%$ depending on the selection criteria) only one merger since $z=1$, in agreement with these authors' formation scenario.
To further shed light on the nuances of the formation pathways followed by our supermassive disks, in this section, we split our sample  more thoroughly into the following four categories:
\begin{itemize}
    \item Galaxies with no significant mergers after $z=1$.
    \item Disk galaxies that suffer a significant merger after $z=1$, reform their disks after the interaction.
    \item Spheroidal galaxies that suffer a significant merger after $z=1$ that consistently had high rotational support.
    \item Spheroidal galaxies that suffer a significant merger after $z=1$, producing a disk as a remnant.
    
\end{itemize}

Figure \ref{fig:path_examples} shows an example of each of the classifications mentioned above, with the morpho-kinematic parameter used to perform the selection displayed as a function of redshift.
Continuous (cyan), dashed (red), dash-dotted (purple), and dotted (brown) lines represent galaxies with a quiescent merging history, reconstructed disks, fast-rotating spheroids, and spheroids to disks, respectively. Colored arrows and stars highlighted when the last significant merger started and finished.
As the dashed (red) line shows, sudden drops in the morpho-kinematic parameter may be interpreted as passages during the merging process. 
As stated previously, in the case of our stricter threshold, 24 galaxies ($48\%$) did not experience any significant interaction after $z=1$. They have a remarkably quiescent merging history and are the most dominant population in our sample, followed by spheroids without significant rotation until the last gas-rich mergers with 13 galaxies ($\sim 26\%$).
Only 10 disk galaxies ($\sim 20\%$) suffered a significant merger at $z < 1$, but were able to reconstruct their disk after the interaction. The remaining 3 ($\sim 6\%$) galaxies were spheroidal before the merger but had significant rotational support even before the merger. 

Our results show that even though the vast majority in our sample did not experience an active merging history, once disks are formed, they are rarely disrupted. Moreover, it only takes one gas-rich merger to trigger a disk in some cases, even though galaxies may be widely dispersion-dominated.

\section{Summary and conclusions}
\label{sec:summary}

In this work, we study the conditions needed for the formation and survival of supermassive disks using the state-of-the-art public database from the magneto-hydrodynamical cosmological simulations of the IllustrisTNG project. 

Using the public database from the high-resolution run from IllustrisTNG100, we use a morpho-kinematic criterion to split our massive galaxies between disks and spheroids as follows:
\begin{itemize}
	\item Galaxies must have a log$_{10}M_\star \geq 11$[M$_\odot$] 
    \item Disks must have a $\lambda / \varepsilon \geq 0.31$ or  0.71.
\end{itemize}
With these criteria, we found 220 and 50 supermassive disks with each morpho-kinematic criterion, respectively, from a total of 694 galaxies with log$_{10}M_\star \geq 11$[M$_\odot$].

The fraction of supermassive disks steadily decreases with stellar mass, but we still find four galaxies with log$_{10}M_\star \geq 11.4$[M$_\odot$]; one having a log$_{10}M_\star \sim 11.8$[M$_\odot$] found regardless of the kinematic threshold.
These results suggest that, as reported in \citet{Quilley22}, the density of disks suddenly drops when a certain stellar mass is reached, in our case being log$_{10}M_\star \geq 11$[M$_\odot$]. Although this does not mean that supermassive disks can't exist, they become less likely depending on the definition.

The distribution of supermassive disks is generally less massive than that of our sample of spheroids. However, even when comparing galaxies within the same stellar mass distribution, as for our control sample, apparent differences arise in their assembly histories.

Supermassive disks tend to have a quiescent merging history, with most of their stellar mass assembled by star formation. In contrast, spheroidal galaxies have a relatively large number of mergers at any stellar mass. Additionally, the few mergers that disk galaxies have tend to be gas-rich, regardless of whether the merger was minor or major.
Supermassive disks can be found in any environment, from isolation to clusters, although they are more prominent in low-density environments. Additionally, we found that they may exist and survive regardless of their condition as satellites or centrals of their host haloes.
When selecting supermassive disk galaxies at $z=0.5$, we found that once those disks become massive enough, the chances of surviving and maintaining their disk morphology are relatively high ($\sim 60\%$) at any selection threshold.
When comparing the impact that AGN may have on the formation or stability of supermassive disks, we found that although some differences may be seen between spheroidal and disk galaxies in the energy injected thermally via AGN, minor differences are seen in the kinetic energy injected into the medium at any stellar mass or kinematic threshold. This result suggests that although AGN plays a key role in regulating the star formation of galaxies, it does not directly change their morphology. 

These results conclude that the number of mergers defines galaxies' morphology and/or kinematical support at any stellar mass.
That said, we found that supermassive disks are long-lived objects with a quiescent merging history that can be more easily found in low-density environments. However, they can reside from isolation to clusters.

These objects become exciting candidates for studying secular galaxy evolution. As this work establishes, their merging history should have been very quiescent in the last 8Gyrs. Studying these objects in the future with upcoming surveys will be key to better understanding the star formation history of rotationally supported supermassive galaxies.

\begin{acknowledgements}
      We want to thank the referee for their helpful and insightful comments, which significantly improved this work. We would also like to thank Yetli Rosas-Guevara, Yara Jaff\'e, and Jacob Crosset for valuable comments and discussions. 
      We gratefully acknowledge support from the ANID BASAL project FB210003. DP acknowledges financial support from ANID through FONDECYT Postdoctorado Project 3230379. GG thanks Pontificia Universidad Católica de Chile, ESO, France-Chile Laboratory of Astronomy (FCLA), and Laboratoire d’Astrophysique de Marseille (LAM) for their support during a 2024-2025 sabbatical leave. PBT acknowledges partial funding by Fondecyt-ANID 1240465/2024 and Núcleo Milenio ERIS NCN2021\_017. AM acknowledges support from the FONDECYT Regular grant 1212046. FAG acknowledges support from the FONDECYT Regular grant 1211370. PBT, AM and FAG acknowledge funding from the HORIZON-MSCA-2021-SE-01 Research and Innovation Programme under the Marie Sklodowska-Curie grant agreement number 101086388. BTC gratefully acknowledges funding by ANID (Beca Doctorado Nacional, Folio 21232155).
\end{acknowledgements}
\bibliographystyle{aa}
\bibliography{quench_final}

\end{document}